\documentclass[10pt,conference]{IEEEtran}
\usepackage[T1]{fontenc}
\usepackage{url}
\usepackage{booktabs} % For nicer table
\usepackage{multirow, multicol}
\usepackage{empheq}
\usepackage{amsmath,amssymb,amsfonts, amsthm, mathtools} % Math packages
\usepackage{dsfont}
\usepackage{bbm, mathrsfs} % Math fonts
\usepackage{algorithm, algorithmic} % For algorithms
\usepackage{graphicx} % To include graphics
\usepackage{textcomp}
\allowdisplaybreaks
\usepackage{caption}
\makeatletter
\def\blfootnote{\gdef\@thefnmark{}\@footnotetext}
\makeatother

\usepackage{subcaption}
\usepackage{eqparbox}

\newcommand{\secref}[1]{\S\ref{#1}}

\newtheoremstyle{custom-theorem-style}  % <name>
        {3pt}                                               % <space above>
        {3pt}                                               % <space below>
        {\normalfont}                               % <body font>
        {}                                                  % <indent amount}
        {\bfseries\itshape}                 % <theorem head font>
        {\normalfont\bfseries.}         % <punctuation after theorem head>
        {.5em}                                          % <space after theorem head>
        {}                                                  % <theorem head spec (can be left empty, meaning `normal')>
\theoremstyle{custom-theorem-style}

\newtheorem{problem}{Problem}

\DeclarePairedDelimiter\abs{\lvert}{\rvert}

\newcommand{\pomdp}[0]{\mathcal{P}}

\newcommand{\textstate}[1]{\texttt{#1}}
\newcommand{\textset}[1]{\mathbf{#1}}

\newcommand{\textfunc}[1]{\mathcal{#1}}

\newcommand{\prob}[1]{\mathbbm{P}[#1]}

\DeclareMathOperator*{\argmin}{arg\,min}

\newcommand{\hoststate}[2]{S^{h_{#1}}_{#2}}

\newcommand{\procstate}[2]{S^{p_{#1}}_{#2}}

\newcommand{\atthost}[1]{A^{h}_{#1}}
\newcommand{\attproc}[1]{A^{p}_{#1}}

\newcommand{\defhost}[1]{D^{h}_{#1}}
\newcommand{\defproc}[1]{D^{p}_{#1}}

\newcommand{\gobs}[1]{G_{#1}}
\newcommand{\lobshost}[2]{O^{h_{#1}}_{#2}}
\newcommand{\lobsproc}[2]{O^{p_{#1}}_{#2}}

\newcommand{\vgobs}[1]{g_{#1}}
\newcommand{\vlobshost}[2]{o^{h_{#1}}_{#2}}
\newcommand{\vlobsproc}[2]{o^{p_{#1}}_{#2}}

\begin{document}

\title{Learning Intrusion Response Strategies \\ for
OT Systems
}

\author{\IEEEauthorblockN{Duc Huy Le, Rolf Stadler}
\IEEEauthorblockA{Dept. of Network and Systems Engineering \\ KTH Royal Institute of Technology, Stockholm, Sweden \\
Email: \{dhle, stadler\}@kth.se} \\

% \today
}

\maketitle

%Comment the below two lines to remove page numbers
% \thispagestyle{plain}
% \pagestyle{plain}

\begin{abstract}
Cyberattacks against Operational Technology (OT) systems, which monitor and control industrial processes, pose an increasing threat to essential societal services. For this reason, developing automated intrusion response strategies is highly important. In this paper, we present a formal model of an OT intrusion response use case using the POMDP framework. It includes a realistic model of partial observability that is based on traffic measurements. This approach allows us to develop tractable, learning-based solution methods for automated intrusion response, which are based on PPO. We evaluate the obtained response strategies on an emulated OT system and find that they are effective against several types of MITRE attacks for the studied use case.
\end{abstract}


%% --------------- INTRODUCTION ------------------------------------
%% -----------------------------------------------------------------

\section{Introduction}

Operational Technology (OT) systems, which monitor and control industrial processes, are increasingly exposed to cyber threats that target critical infrastructure and industrial production. The number of OT cyberattacks grew by more than $90\%$ annually between 2019 and 2023~\cite{waterfall-report}. Ransomware attacks against industrial organizations increased by $87\%$ in 2024~\cite{dragos-report}. Although the number of OT attacks is currently lower than that of Information Technology (IT) attacks, their consequences can be more severe, by disrupting essential services and affecting communities beyond the targeted organizations~\cite{waterfall-report}. Notable examples of such attacks include TRITON/TRISIS~\cite{incident-triton}, Industroyer2~\cite{incident-industroyer}, and FrostyGoop~\cite{dragos-report}.

%Operational Technology (OT) systems are increasingly exposed to cyber threats that target at critical infrastructure and industrial production. Recent reports indicate a substantial increase in OT cybersecurity incidents. The number of reported OT cyberattacks grew at more than $90\%$ annually between 2019 and 2023~\cite{waterfall-report}, while ransomware attacks against industrial organizations increased by $87\%$ in 2024~\cite{dragos-report}. Although OT incidents remain fewer than Information Technology(IT) cyberattacks, their consequences can be disproportionately severe, potentially disrupting essential services and affecting communities beyond the targeted organization~\cite{waterfall-report}. Notable examples include TRITON/TRISIS targeting at a petrochemical facility in Saudi Arabia~\cite{incident-triton}, the Industroyer2 attacks against Ukrainian power-grid operations~\cite{incident-industroyer}, the Unitronics PLC campaign affecting water systems in the United States, Ireland, and other countries~\cite{dragos-report}, and the FrostyGoop attacks against district-heating infrastructure in Ukraine~\cite{dragos-report}.

A factor contributing to the rise of OT attacks is the IT/OT convergence, where IT systems are integrated with physical OT networks to facilitate monitoring, data collection, and efficient operation. This integration expands the attack surface and creates paths from enterprise networks into OT environments, which adversaries can exploit~\cite{it-ot-convergence1, it-ot-convergence2}. Traditionally, OT infrastructures have a lower level of cybersecurity than IT infrastructures~\cite{ot-cybersec-weakness}.

%A major factor contributing to the rise in OT attacks is the IT/OT convergence, where IT systems are integrated with physical OT networks to support centralized monitoring, data collection, and operational efficiency. However, it also expands the attack surface by creating paths from enterprise networks into OT environments that adversaries can exploit~\cite{it-ot-convergence1, it-ot-convergence2}. This exposure is further amplified by persistent security weaknesses in OT infrastructure, including weak authentication, unpatched firmware, exploitable legacy devices, and insecure remote access~\cite{ot-cybersec-weakness}.

OT cybersecurity guidelines, e.g. \cite{ics-security-book, nist-ot-security, isa-iec-guideline}, emphasize proactive measures, such as network segmentation  and vulnerability analysis, as well as monitoring-based control, including resetting hosts and isolating network segments. They leave intrusion response actions largely to human operators and do not stress automated responses.

%Existing OT cybersecurity guidance, e.g. \cite{ics-security-book, nist-ot-security, isa-iec-guideline}, primarily emphasizes preventive and monitoring-oriented controls, such as secure architecture, network segmentation, asset visibility, vulnerability management, and continuous monitoring, leaving intrusion response largely to human operators once an attack is detected. This is problematic in OT environments, where attacks are time-critical and delayed response can allow attacker progression and significantly disrupt system operations. 

Early research on automated intrusion response for IT and OT systems has focused on rule-based policies, which configure firewall rules~\cite{related-work-rule-based} or SDN flows~\cite{related-work-rule-based2, related-work-rule-based3}, for instance. These rules are defined and maintained by human experts.

% SHORTEN VERSION:
% Recent work has studied learning-based intrusion response for OT systems, mainly using Reinforcement Learning (RL). Existing methods either learn heuristic policies without a formal model~\cite{related-work-heuristic, related-work-heuristic2, related-work-heuristic3}, or use formal models that assume full observability of the system state or attacker actions ~\cite{related-work-mdp, related-work-mdp2, related-work-mdp3, related-work-game, related-work-game2}. Alternatively, POMDP-based approaches assume specific obervation models, without explaining how such boervations can be obtained in real systems~\cite{related-work-pomdp, related-work-pomdp2}.

Recently, methods for OT intrusion response have been developed that do not rely on predefined rules, but are based on learning from system measurements. These methods use Reinforcement Learning (RL) as the dominant concept. Some RL approaches learn heuristic policies without a formal system model~\cite{related-work-heuristic, related-work-heuristic2, related-work-heuristic3}, which precludes the understanding of an achievable optimal strategy. Other works apply formal models, such as Markov Decision Process (MDP)~\cite{related-work-mdp, related-work-mdp2, related-work-mdp3}, Partially Observable Markov Decision Process (POMDP)~\cite{related-work-pomdp, related-work-pomdp2}, and stochastic game~\cite{related-work-game, related-work-game2}. 
Most of these lines of research assume full observability of the system state or attacker actions, which is unrealistic and limits the practical relevance of the results. Alternatively, some studies assume specific observation models, without detailing how such observations can be obtained in real systems~\cite{related-work-pomdp, related-work-pomdp2}.

\begin{figure}[h!]
    \centering
    \includegraphics[width=0.92\linewidth]{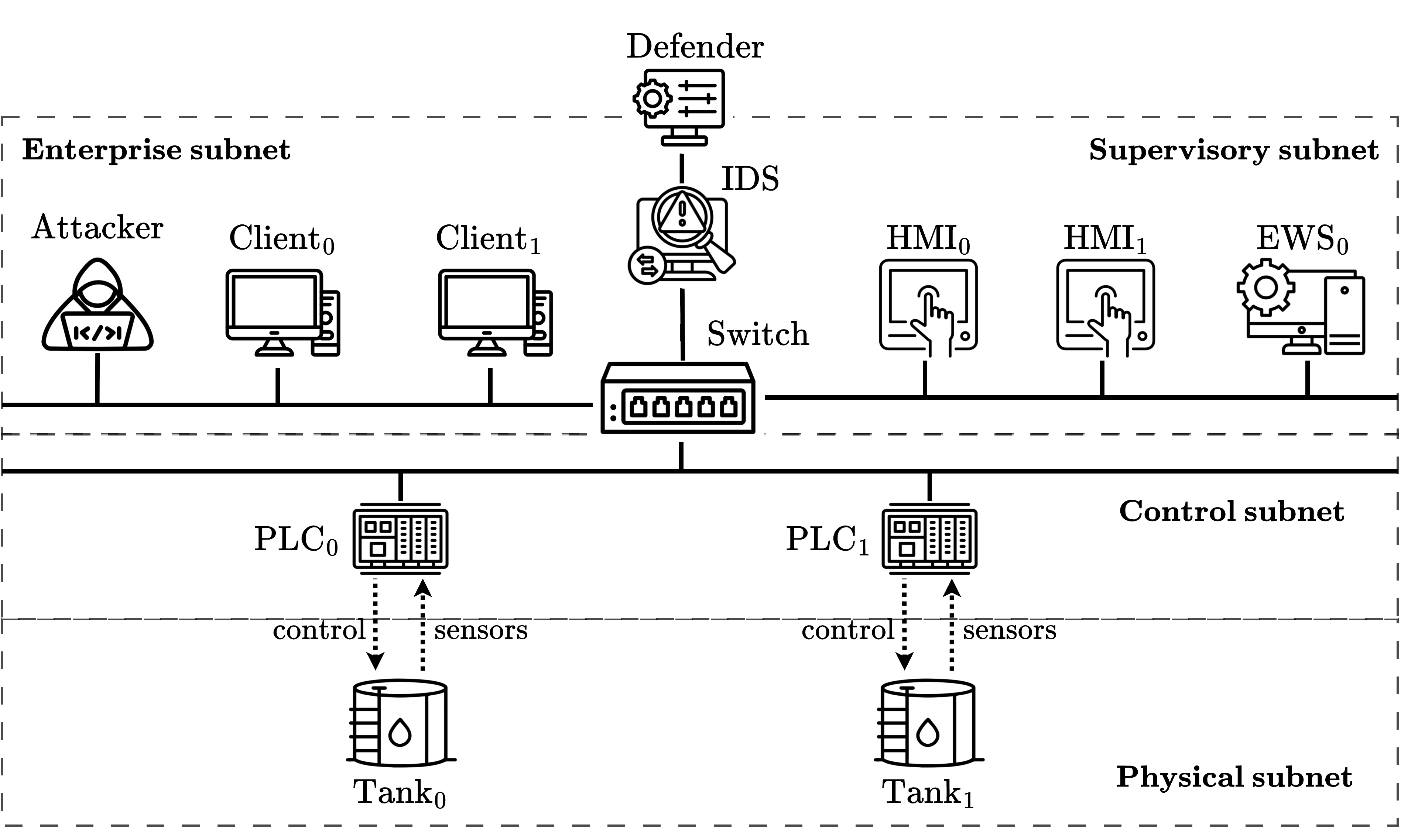}
    \caption{The OT infrastructure of the intrusion response use case}
    \label{fig:ot-topo}
\end{figure}

In this paper, we investigate an intrusion response use case for an OT infrastructure (see Fig. \ref{fig:ot-topo}). The infrastructure follows the Purdue model architecture with network segmentation~\cite{purdue-model}. The objective of the attacker is to disrupt the industrial processes that control the tanks in the OT infrastructure. The attacker creates a path through the infrastructure by executing a sequence of actions, including reconnaissance, vulnerability scanning, exploitation, and host inspection. The defender's objective is to maintain normal process operation and mitigate the attacker actions. The defender monitors the infrastructure and executes a sequence of defensive actions based on continuous observations of the network traffic.

We model this scenario as a discrete-time dynamical system and formalize the defender’s problem as a Partially Observable Markov Decision Process (POMDP)~\cite{pomdp}. The model allows us to capture attacker and defender actions and define an optimal defender strategy for the use case. To learn effective defender strategies, we propose two learning methods, which are based on Proximal Policy Optimization (PPO), a state-of-the-art RL algorithm. Applying PPO to our problem formulation is not computationally feasible, since the state space (and thus the belief space) is too large. The first method, $k$-Obs-PPO, simplifies the structure of the input for the defender strategy, while the second method, BF-PPO, uses particle filter~\cite{particle_filter} to estimate the system state. 

We have developed an emulation system, which serves as a replica of the target infrastructure shown in Fig.~\ref{fig:ot-topo}. On this system, we perform the attacks for this scenario. While the attacks are occurring, we collect traffic measurements. These measurements are used to estimate the observation function of the POMDP model, which allows us to simulate this model and to compute effective defender strategies. The obtained defender strategies are evaluated on the emulation system.

The key results of this investigation are as follows. We can compute effective defender strategies for the selected scenario using our methods. They converge in reasonable time on a simulator (about $1$ hour on an Apple M3 Pro processor). BF-PPO produces a defender strategy that performs on the emulation system close to a strategy that assumes full observability, for three different types of attacks. 

We make four contributions with this paper:

\begin{itemize}
    \item We present a formal model of the OT intrusion response use case using the POMDP framework and formally define an optimal defender strategy.
    \item We present two solution methods, $k$-Obs-PPO and BF-PPO, for automated response. The methods are learning based, not rule-based, and not set up by experts. They efficiently learn effective defender strategies.
    \item Our POMDP model includes a realistic model of partial observability, which is based on online traffic measurements. 
    \item We evaluate the solution methods in a realistic emulation environment of the OT infrastructure, demonstrating that the learned defender strategies are effective. In particular, BF-PPO achieves performance close to a defender strategy that has been computed under the assumption of full observability.
\end{itemize}

We see the main limitations of this work as follows. First, like all formal models, our model relies on simplifying assumptions. For instance, we assume that attacker and defender actions are taken simultaneously at discrete times and have immediate effects. In reality, such actions occur asynchronously in continuous time, have different durations, and affect the system with delays. Capturing such dynamics would make the model more realistic, at the expense of increasing the computational complexity of the learning defender strategies. Also, we could have modeled the state space with higher granularity, with the same drawback. Second, our POMDP model is valid for the specific use case, system configuration and attack types. In this work, we did not study the generalization property of our model with respect to those aspects. Third, this work currently does not consider operational safety, which constrains defender strategy. We will do so in the future.

%  Modeling of reality; How accurately does our model and assumptions capture the reality of an OT system?
% There is a tradeoff between accurate modeling and computational complexity and sample efficiency of the solution method

% Generalization: other attacks, other system configurations, other background

%% --------------- RELATED WORKS -----------------------------------
%% -----------------------------------------------------------------

\section{Related Work}
\label{sec:related-work}

Many studies have investigated the intrusion response problem in IT systems. They use a variety of modeling frameworks, including POMDP~\cite{related-work-it-pomdp, related-work-it-pomdp2}, game theory~\cite{related-work-it-game, related-work-it-game2}, or causal models ~\cite{related-work-it-causal-model}. They apply numerous solution techniques, such as RL~\cite{related-work-it-pomdp, related-work-it-pomdp2}, attack-graph-based planning~\cite{related-work-it-attack-graph}, and large language models~\cite{related-work-it-llm}.

Automated intrusion response for OT systems has been studied using rule-based and learning-based approaches. Rule-based methods define and enforce predefined response actions, for example through automated firewall configuration~\cite{related-work-rule-based} or software-defined networking~\cite{related-work-rule-based2, related-work-rule-based3}. These methods automatically detect possible incidents and their response strategies are defined by human experts.

Recent advances in RL have initiated research on learning-based intrusion response solutions for OT systems, which allows defender strategies to be learned from measurements. Some studies propose heuristic solution methods that do not require a formal system model of the use case~\cite{related-work-heuristic, related-work-heuristic2, related-work-heuristic3, related-work-heuristic4}. However, without a formal model, it cannot be assessed to which extent the learned defender strategies are optimal. 

Other works formulate intrusion response problems using decision-theoretic frameworks, including MDP~\cite{related-work-mdp,related-work-mdp2,related-work-mdp3}, POMDP~\cite{related-work-pomdp,related-work-pomdp2}, and stochastic games~\cite{related-work-game, related-work-game2}. Most of them investigate scenarios in which the defender can observe the attacker’s actions and progress. This assumption of full observability is unrealistic. Some works consider partial observability using abstract observation models, but do not discuss how such models are obtained for real systems~\cite{related-work-heuristic3, related-work-pomdp, related-work-pomdp2}. A methodology for developing a defender strategy, which first computes the strategy based on a formal model and subsequently evaluates it on an emulation platform, is presented in~\cite{kim-thesis}. 

The work in this paper is unique in that it considers the following aspects in combination. It formulates a formal model of an OT intrusion response use case, presents learning-based solution methods for automated intrusion response, includes a realistic model of partial observability and evaluation results from an emulated OT system.

%% --------------- CAGE-2 -----------------------------------
%% -----------------------------------------------------------------

\section{The OT Intrusion Response Use Case}
\label{sec:use-case}

We study an intrusion response use case for the OT infrastructure shown in Fig.~\ref{fig:ot-topo}. The infrastructure follows the Purdue model~\cite{purdue-model} and is partitioned into enterprise, supervisory, control, and physical subnets. The \textit{enterprise} subnet contains client systems that provide access to the industrial environment. The \textit{supervisory} subnet contains servers with Human--Machine Interfaces (HMIs) and Engineering Workstations (EWSs) for process monitoring, control, and engineering operations. The \textit{control} subnet contains Programmable Logic Controllers (PLCs), which manage two water tanks in the \textit{physical} subnet.

% We study an intrusion-response use case in a water storage system, a small-scale OT infrastructure where water storage units are monitored and controlled by programmable control devices. The system is designed according to the Purdue model~\cite{purdue-model}, a widely accepted reference architecture for security-aware ICS and OT network segmentation~\cite{ics-security-book,nist-ot-security}. The infrastructure is partitioned into four logical subnets, including enterprise, supervisory, control, and physical (see Fig. \ref{fig:ot-topo}). The \textit{enterprise} subnet hosts client systems that represent business-side users and remote access points to the industrial environment. The \textit{supervisory} subnet contains Human--Machine Interfaces (HMIs) and an Engineering Workstation (EWS). The HMIs support process monitoring and operator interaction, whereas the EWS supports engineering activities such as monitoring, configuration, diagnostics, and maintenance. The \textit{control} subnet contains Programmable Logic Controllers (PLCs), which execute control logic to maintain the water levels of the tanks at desired setpoints. Finally, the \textit{physical} subnet contains the two water tanks together with their associated sensors and actuators.

We assume the attacker has a foothold in the enterprise subnet, either through a compromised enterprise host or through insider access to the IT network. The attacker’s objective is to disrupt the industrial processes by tampering with the tanks. We consider attacks from the MITRE ATT\&CK framework~\cite{mitre}, where an attacker performs reconnaissance, exploitation, and inspection on supervisory hosts, and then uses a compromised host to interact with the PLCs controlling the tanks. We consider three attacker strategies for this use case, which differ in the way they perform reconnaissance, exploitation, etc.

% The attacker is assumed to have established an initial foothold in the enterprise subnet, either by compromising a client or through lateral movement within the IT network. Its goal is to maliciously tamper the operation of the physical processes. To achieve this goal, the attacker is equipped with standard techniques described in the MITRE ATT\&CK for ICS framework \cite{mitre}. These techniques include: (1) scanning the supervisory subnet to discover active hosts, (2) scanning a supervisory host to identify exposed services and vulnerabilities, (3) exploiting a host to gain access, (4) inspecting a compromised host to infer its control relationship with the physical processes, and (5) tampering with a physical process through the compromised host.

The defender cannot directly observe the attacker’s progression through the infrastructure. Instead, it receives network measurements from the switch through the IDS in Fig.~\ref{fig:ot-topo}. The defender’s objective is to maintain normal operation of the industrial processes and mitigate the attacker actions. The defender can perform three types of defensive actions: reset a supervisory host, reset an industrial process, and reset all hosts in the supervisory subnet and the control subnet. A reset action on a component reboots the machine, renews its credentials, changes its IP address, etc. Such an action can cause a temporary disruption of industrial operations.

Note that the traffic measurements not only relate to the attacker actions but also to normal process operations, which makes their interpretation for the defender difficult.

%% --------------- POMDP Formalisation ----------------------------
%% ----------------------------------------------------------------

\section{Formalizing the Use Case Using a POMDP Model}
\label{sec:pomdp}

We study the intrusion response use case introduced in \secref{sec:use-case} from the defender’s perspective. We assume that the system evolves in discrete time steps with a finite horizon. At each time, both the defender and the attacker take an action. The defender cannot observe the attacker’s progression and obtains knowledge about the attack through network measurements. We formalize the problem of intrusion response as a sequential decision-making problem under partial observability. We choose POMDP as the modeling framework.

%%%%%%%%%%%%%%%%%%%%%% Theoretical background %%%%%%%%%%%%%%%%%%%%%%%%%%%%%
%%%%%%%%%%%%%%%%%%%%%% Theoretical background %%%%%%%%%%%%%%%%%%%%%%%%%%%%%
%%%%%%%%%%%%%%%%%%%%%% Theoretical background %%%%%%%%%%%%%%%%%%%%%%%%%%%%%

\subsection{Partially Observable Markov Decision Process}

A Partially Observable Markov Decision Process (POMDP) models the decision process of a discrete-time Markovian system with partial observability~\cite{pomdp}. It is defined by a 10-tuple $\pomdp = (\textset{S}, \textset{D}, \textfunc{T}, \textset{O}, \mathcal{Z}, \mathcal{C}, \gamma, \rho_1, T, \textset{B})$. $\textset{S}$ denotes the state space and $\textset{D}$ denotes the action space. The probability measure $\textfunc{T}: \textset{S}\times\textset{S}\times \textset{D} \to [0,1]$ denotes the system dynamics, where $\textfunc{T}(S_{t+1}|S_t,D_t)$ is the transition probability from state $S_t$ to $S_{t+1}$ by taking action $D_t$. The state transition is partially observable through variable $O_t \in \textset{O}$, where $\textset{O}$ is the observation space. The conditional observation distribution is denoted as $\mathcal{Z}$. Taking an action $D_t$ in state $S_t$ induces a cost $C_t =\textfunc{C}(S_t, D_t) \in \mathds{R}$. The objective in a POMDP is to find the sequence of $T$ actions, $D_1, \dots, D_T$, to minimize the expected cumulative cost $\mathds{E}[J]$, with discount factor $\gamma \in (0,1]$:

\begin{equation}
    J = \sum_{t=1}^T \gamma^{t-1} C_t
\end{equation}

A belief state $b_t = \langle b_t(s_t) \rangle_{s_t \in \textset{S}}$ is associated with time $t$, where $b_t(s) = \prob{S_t = s | h_t}$ with $h_t = (\rho_1,o_1,d_1,o_2,\dots, d_{t-1}, o_t) \in \mathcal{H}$. The belief state is a distribution over the state space $\textset{S}$. At every time $t$, the belief is recursively computed:

\begin{equation}
\label{eq:belief-update}
b_{t+1}(s) = \eta^{-1} \textfunc{Z}(o_{t+1}|s,d_{t}) \sum_{s_t \in \textset{S}} \textfunc{T}(s|s_t,d_t) b_t(s_t) 
\end{equation}

\noindent where $\eta = \sum_{s \in \textset{S}} {\textfunc{Z}(o_{t+1}|s,d_{t}) \sum_{s_t \in \textset{S}} \textfunc{T}(s|s_t,d_t) b_t(s_t)}$ is the normalization factor. At the beginning of an episode, the initial state $S_1$ is sampled from the initial state distribution $\rho_1 : \textset{S} \to [0,1]$, which also defines the initial belief state.

%%%%%%%%%%%%%%%%%%%%%% Formalization %%%%%%%%%%%%%%%%%%%%%%%%%%%%%
%%%%%%%%%%%%%%%%%%%%%% Formalization %%%%%%%%%%%%%%%%%%%%%%%%%%%%%
%%%%%%%%%%%%%%%%%%%%%% Formalization %%%%%%%%%%%%%%%%%%%%%%%%%%%%%

\subsection{A POMDP Model of the Intrusion Response Use Case}

This subsection defines the POMDP model of the defender’s problem. The system state captures attacker progression and process integrity, while the observation represents IDS-derived network measurements. The transition function models the interaction between attacker and defender actions, and the cost function encodes the defender’s objective. The resulting POMDP defines the optimization problem for computing an optimal defender strategy.

\subsubsection{Scenario configuration}

To simplify notation, we denote the supervisory hosts and the processes in Fig. \ref{fig:ot-topo} by
$h_0=\mathrm{HMI}_0$, $h_1=\mathrm{HMI}_1$, and $h_2=\mathrm{EWS}_0$, $p_0=\mathrm{Tank}_0$ and $p_1=\mathrm{Tank}_1$. Let $\textset{H} = \{h_0,h_1,h_2\}$ be the set of hosts and $\textset{P} = \{ p_0, p_1\}$ be the set of tanks. In this topology, each supervisory host $h \in \textset{H}$ can monitor and control both $p_0$ and $p_1$ through their corresponding control units, $\mathrm{PLC}_0$ and $\mathrm{PLC}_1$. During an episode, the attacker follows a fixed strategy $\pi_A \in \Pi_A$, where $\Pi_A$ is the attacker strategy space.

\subsubsection{System state space $\textset{S}$ and initial state distribution $\rho_1$}

The system state $S_t$ at time $t$ represents the states of the supervisory hosts, the states of the industrial processes, and the attacker action state from the previous time step. Formally, $S_t$ is defined as the tuple $S_t=(\hoststate{0}{t},\hoststate{1}{t}, \hoststate{2}{t}, \procstate{0}{t},\procstate{1}{t},A_{t-1})$.

For each host $h \in \textset{H}$, its state $\hoststate{}{t} \in \textset{S^{h}} = \{\textstate{U}, \textstate{D}, \textstate{S}, \textstate{E}, \textstate{I}\}$ represents the attacker’s progression on host $h$ at time $t$. The values denote whether $h$ is undiscovered by the attacker ($\textstate{U}$), discovered by the attacker ($\textstate{D}$), scanned by the attacker ($\textstate{S}$), exploited and accessed by the attacker ($\textstate{E}$), or inspected for PLC control by the attacker ($\textstate{I}$). These states are motivated by MITRE ATT\&CK framework~\cite{mitre} and reflect the main attack stages of an OT cyberattack, where an attacker first discovers assets, then gathers information, gains access, and learns control patterns.

% For each supervisory host $h \in \textset{H}$, the state $\hoststate{}{t}$ represents the attacker’s progression on host $h$ at time $t$. It takes one of the following values: $\textstate{U}$ if the host is undiscovered by the attacker, $\textstate{D}$ if the host is discovered by the attacker, $\textstate{S}$ if the host has been scanned by the attacker, $\textstate{E}$ if the attacker has successfully exploited and gained access to the host, and $\textstate{I}$ if the control of the processes has been inspected by the attacker. The state space of $\hoststate{}{t}$ is $\textset{S^{h}} = \{\textstate{U}, \textstate{D}, \textstate{S}, \textstate{E}, \textstate{I}\}$.

% The states of a supervisory host reflect the main attack stages of an OT cyber intrusion. They are motivated by MITRE ATT\&CK framework~\cite{mitre}, where an attacker first discovers assets, then gathers information, gains access, and learns control pattern. Thus, the states $\textset{S^{h}}$ provide a compact abstraction of the main stages of attacker progression before process-level impact. 

For each process $p \in \textset{P}$, its state $\procstate{}{t} \in \textset{S^{p}} = \{\textstate{W}, \textstate{C}\}$ denotes the operation condition of the water tanks. The values denote whether $p$ is in its normal operation state ($\textstate{W}$) or has been corrupted by the attacker ($\textstate{C}$).

% For each process $p \in \textset{P}$, the state $\procstate{}{t}$ represent the the security state of process $p$ at time $t$. It takes one of the two values: $\textstate{W}$ if the process is in normal operation state, and $\textstate{C}$ if the process is corrupted, i.e., tampered by the attacker. The state space of $\procstate{}{t}$ is $\textset{S^{p}} = \{\textstate{W}, \textstate{C}\}$.

The component $A_{t-1}$ denotes the attacker action at time $t-1$ and is defined as $A_{t-1} = (\atthost{t-1}, \attproc{t-1})$. $\atthost{t-1}$ represents the targeted host at time $t-1$ and takes three types of values: $\emptyset$, if no host is targeted; $h \in \textset{H}$, if host $h$ is targeted; and $\textset{H}$, if the attacker scans the supervisory subnet. $\attproc{t-1}$ represents the targeted process at time $t-1$ and can be either $\emptyset$ or a process $p \in \textset{P}$. We require $\attproc{t-1} \neq \emptyset$ only when $\atthost{t-1} \in \textset{H}$, meaning that process tampering is performed through a targeted host. The attacker action space is $\textset{A} = \{\textset{H} \cup \{\emptyset, \textset{H}\}\} \times \{\textset{P} \cup \{\emptyset\}\}$.

% $A_{t-1}$ denotes the attacker action at time $t-1$ and is represented by two components, $A_{t-1} = (\atthost{t-1}, \attproc{t-1})$. The component $\atthost{t-1}$ represents the host target of the attacker at time $t-1$ and can take three types of values: $\emptyset$, if no host is targeted; $h \in \textset{H}$, if a specific host $h$ is targeted; and $\textset{H}$, if the attacker scans the supervisory subnet. The component $\attproc{t-1}$ represents the process target of the attacker at time $t-1$ and can be either $\emptyset$ or a process $p \in \textset{P}$. We require $\attproc{t-1} \neq \emptyset$ only when $\atthost{t-1} \in \textset{H}$, meaning that process tampering is performed through a targeted supervisory host. The attacker action space is $\textset{A} = \{\textset{H} \cap \{\emptyset\}\} \times \{\textset{P} \cup \{\emptyset\}\}$.

The state space is defined as $\textset{S} = \textset{S^{h}}^3 \times \textset{S^{p}}^2 \times \textset{A}$. At time $t=1$, no intrusion has occurred, hence, the initial state distribution $\rho_1$ is a degenerate distribution with $\rho_1(s)=1$, where $s = (\textstate{U}, \textstate{U}, \textstate{U}, \textstate{W}, \textstate{W}, (\emptyset, \emptyset))$.

\subsubsection{Defender action space $\textset{D}$}

\bgroup
\begin{figure*}[htbp]
    \centering
    \includegraphics[width=0.85\linewidth]{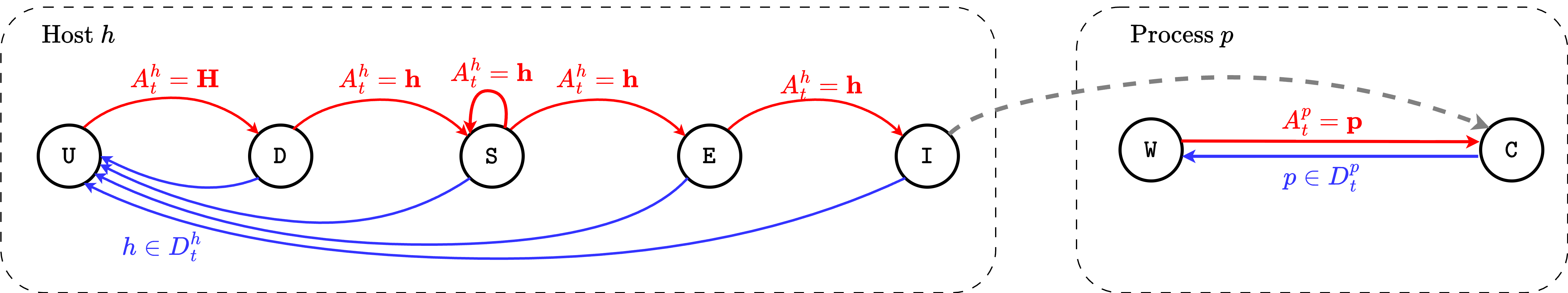}
    \caption{The transitions of host states $\hoststate{}{t}$ and the process states $\procstate{}{t}$ caused by attacker action $A_t$ (red arrows) and defender action $D_t$ (blue arrows). The dashed arrow illustrates that host $h$ must be in state $\textstate{I}$ for the attacker to tamper with process $p$ using host $h$ ($A^h_t =h, A^p_t=p)$.}
    \label{fig:transition_function}
\end{figure*}
\egroup

At time $t$, the defender takes an action $D_t = (\defhost{t},\defproc{t})$. $\defhost{t}$ is the target host(s) and $\defproc{t}$ is the target process(es) at time $t$. $D_t$ can take one of the following values: $(\defhost{t} = \emptyset, \defproc{t} = \emptyset)$ if the defender passively monitors the system; $(\defhost{t} = h \in \textset{H}, \defproc{\emptyset})$ if the defender resets host $h$; $(\defhost{t} = \emptyset, \defproc{t} = p \in \textset{P})$ if the defender resets process $p$; and $(\defhost{t} = \textset{H}, \defproc{t} = \textset{P})$ if the defender resets the whole system. The defender action space is thereby defined as $\textset{D} = \{(\emptyset,\emptyset)\} \cup \{(h,\emptyset) : h \in \textset{H}\} \cup \{(\emptyset,p) : p \in \textset{P}\} \cup \{(\textset{H},\textset{P})\}.$

\subsubsection{Transition function $\textfunc{T}$}

The state transition $\mathcal{T}(S_{t+1}|S_{t},D_t)$ from time $t$ to $t+1$ can be described through the transitions of each component of the system state: (a) $A_{t-1} \rightarrow A_t$; (b) $\hoststate{}{t} \rightarrow \hoststate{}{t+1}$; and (c) $\procstate{}{t} \rightarrow \procstate{}{t+1}$. The transitions are presented as follows.

\paragraph{Attacker action $A_{t-1}\rightarrow A_{t}$}

The attacker action $A_t$ at time $t$ is defined by the system state $S_t$ and its strategy $\pi_A$, i.e., $A_t = \pi_A(S_t)$.

\paragraph{Host state $\hoststate{}{t} \rightarrow \hoststate{}{t+1}$}

The state $\hoststate{}{t+1}$ of a host $h \in \textset{H}$ is dependent on its previous state $\hoststate{}{t}$, the attacker action $A^h_t$ and the defender action $D^h_t$. 
\begin{subequations}
\label{eq:pomp-trans:host-state}
\begin{align}
\label{eq:pomdp-trans:host:discover}\hoststate{}{t} = \textstate{U} &\to \hoststate{}{t+1} = \textstate{D} \text{ if } A^h_t = \textset{H}, h \notin D^h_t \\
\label{eq:pomdp-trans:host:scan}\hoststate{}{t} = \textstate{D} &\to \hoststate{}{t+1} = \textstate{S} \text{ if } A^h_t = \textset{h}, h \notin D^h_t \\
\label{eq:pomdp-trans:host:exploit-success}\hoststate{}{t} = \textstate{S} &\to \hoststate{}{t+1} = \textstate{E} \text{ if } A^h_t = \textset{h}, N_{e} = 1, h \notin D^h_t \\
\label{eq:pomdp-trans:host:exploit-fail}\hoststate{}{t} = \textstate{S} &\to \hoststate{}{t+1} = \textstate{S} \text{ if } A^h_t = \textset{h}, N_{e} = 0, h \notin D^h_t \\
\label{eq:pomdp-trans:host:inspect}\hoststate{}{t} = \textstate{E} &\to \hoststate{}{t+1} = \textstate{I} \text{ if } A^h_t = \textset{h}, h \notin D^h_t \\
\label{eq:pomdp-trans:host:reset}\hoststate{}{t} &\to \hoststate{}{t+1} = \textstate{U} \text{ if } h \in D^h_t \\
\label{eq:pomdp-trans:host:otherwise}\hoststate{}{t} &\to \hoststate{}{t+1} = \hoststate{}{t} \text{ otherwise }
\end{align}
\end{subequations}

\noindent where $N_e$ is a binary random variable that determines the success of the attacker's \textit{Exploit} action on host $h$. 

(\ref{eq:pomdp-trans:host:discover}-\ref{eq:pomdp-trans:host:otherwise}) describe the transition of the host state $\hoststate{}{t} \rightarrow \hoststate{}{t+1}$, $\forall h \in \textset{H}$. \eqref{eq:pomdp-trans:host:discover} describes the attacker's network reconnaissance action on the supervisory subnet, which exposes the hosts $\textset{H}$ to the attacker and causes the transition $\textstate{U} \to \textstate{D}$. \eqref{eq:pomdp-trans:host:scan} describes the transition $\textstate{D} \to \textstate{S}$, where the attacker performs a host-level discovery on host $h$. \eqref{eq:pomdp-trans:host:exploit-success} and \eqref{eq:pomdp-trans:host:exploit-fail} describe the exploitation attempt of the attacker on host $h$, which has a success probability defined by a binary random variable $N_{e}$. After gaining access to host $h$, the attacker inspects it to learn control patterns to the underlying physical processes, which is presented by the transition $\textstate{E} \to \textstate{I}$. All transitions in (\ref{eq:pomdp-trans:host:discover}-\ref{eq:pomdp-trans:host:inspect}) occur under the condition that the defender does not reset host $h$, which brings $\hoststate{}{t+1}$ to the secure state $\textstate{U}$ (\ref{eq:pomdp-trans:host:reset}). 

The transitions are illustrated in Fig.~\ref{fig:transition_function}, which describes the attacker actions on a host toward state $\textstate{I}$. The defender can reset the host and bring it to the secure state $\textstate{U}$.

\paragraph{Process state $\procstate{}{t} \rightarrow \procstate{}{t+1}$}

The state $\procstate{}{t+1}$ of a process $p \in \textset{P}$ is dependent on its state $S_t$, the attacker action $A^p_t$ and the defender action $D^p_t$. The transition is illustrated in Fig. \ref{fig:transition_function}.
\begin{subequations}
\label{eq:pomp-trans:proc-state}
\begin{align}
\label{eq:pomdp-trans:proc:tamper}\procstate{}{t} \to & \procstate{}{t+1} = \textstate{C} \text{ if } A^p_t = p, A^h_t=h, \hoststate{}{t}=\textstate{I}, p \notin D^p_t \\
\label{eq:pomdp-trans:proc:reset}\procstate{}{t} \to  & \procstate{}{t+1} = \textstate{W} \text{ if } p \in D^p_t \\
\label{eq:pomdp-trans:proc:otherwise}\procstate{}{t} \to & \procstate{}{t+1} = \procstate{}{t} \text{ otherwise }
\end{align}
\end{subequations}

(\ref{eq:pomdp-trans:proc:tamper}-\ref{eq:pomdp-trans:proc:otherwise}) describe the transition of the process state $\procstate{}{t} \to \procstate{}{t+1}$, $\forall p \in \textset{P}$. (\ref{eq:pomdp-trans:proc:tamper}) describes the process tampering action, which requires an inspected host ($\exists\;h\in \textset{H}: \hoststate{}{t}=\textstate{I}$), and the process is not reset by the defender at the same time ($p \notin D^p_t$). (\ref{eq:pomdp-trans:proc:reset}) describes the defender resetting process $p$ to its normal working state.

% \subsubsection{Time horizon $T$}
% We model the scenario has a finite horizon $T$

\subsubsection{Defender observation space $\textset{O}$}

The defender cannot observe the system state $S_t$. Instead, the defender observes 
\begin{equation}
\label{eq:pomdp-observation}
O_t = (\gobs{t}, \lobshost{0}{t}, \lobshost{1}{t}, \lobshost{2}{t} ,\lobsproc{0}{t}, \lobsproc{1}{t})
\end{equation}

\noindent where $\gobs{t}$ represents the number of IP packets between the subnets, $\lobshost{i}{t}$ represents the number of IP packets between supervisory host $i$ and the enterprise subnet, and $\lobsproc{j}{t}$ represents the number of IP packets between $\mathrm{PLC}_j$ and the supervisory subnet, during time step $t$.

\subsubsection{Defender observation function $\textfunc{Z}$}

The observation function is the conditional probability mass function of the discrete observation vector $O_{t+1}$:
\begin{equation}
\label{eq:pomdp-observation-function}
\textfunc{Z}(o,s,d) = \prob{O_{t+1}=o \mid S_{t+1}=s, D_{t=d}}
\end{equation}

\noindent where $s \in \textset{S}$, $d \in \textset{D}$, and $o \in \textset{O}$.

\subsubsection{Cost function $\textfunc{C}$ and defender objective $J$}

At each time $t$, the defender pays a cost $C_t$:

\begin{equation}
\label{eq:pomdp:reward}
    C_t = \sum_{h \in \textset{H}} \sigma_h (\hoststate{}{t}) +
    \sum_{p \in \textset{P}} \sigma_p (\procstate{}{t}) +
    \sigma_d (D_t)
\end{equation}

\noindent where $\sigma_h, \sigma_p$, and $\sigma_d$ are parameter functions, which reflect the defender's objective by denoting costs for the host states, the process states and the defensive actions. The functions are specified in Tab.~\ref{tab:cost-parameter}. The high cost for process corruption, $\sigma_p(\textstate{C})=10$, prioritizes the safety of the physical processes. The increasing host state costs $\sigma_h$ penalize the defender for attacker progression and encourage preventive responses. The action cost $\sigma_d$ penalizes resets, preventing unnecessarily disruptive responses.

% \noindent where $\sigma_h : \textset{S}^h \to \mathds{R}$ is a function defining the cost for a host state; $\sigma_p : \textset{S}^p \to \mathds{R}$ is a function defining the cost for a process state; and $\sigma_D : \textset{D} \to \mathds{R}$ is a function defining the cost for a defender action. The mapping of these parameters is specified in Tab.~\ref{tab:cost-parameter}, reflecting the defender’s objective. The high cost assigned to process corruption ($\sigma_p(\textstate{C}) = 10$) drives the defender to prioritize maintaining normal process operation. The host-state costs $\sigma_h$ increase with the attacker’s progression, creating pressure for preventive responses against attacker's progression. Finally, resetting hosts or processes also incurs a cost, preventing the defender from applying disruptive actions unnecessarily.

\bgroup
\def\arraystretch{1.1} %Change margin inside cells
\begin{table}[ht]
    \centering 
    \caption{Parameters of the cost function $\textfunc{C}$}
    \tiny
    \begin{tabular}{cp{0.35\textwidth}}
    \toprule
        \textbf{Cost function} & \textbf{Value}  \\
        \midrule
        $\sigma_h$ & $\sigma_h(\textstate{U}) = \sigma_h(\textstate{D}) = 0, \sigma_h(\textstate{S}) = 0.2, \sigma_h(\textstate{E}) = 1.5, \sigma_h(\textstate{I})=3.5$ \\
        \midrule
        $\sigma_p$ & $\sigma_p(\textstate{W}) = 0, \sigma_p(\textstate{C}) = 10$ \\
        \midrule
        $\sigma_d$ & $\sigma_d(D^h_t, D^p_t) = 2 \times \abs{D^h_t} + 3 \times \abs{D^p_t} $ \\
    \bottomrule
    \end{tabular}
    \label{tab:cost-parameter} 
\end{table}
\egroup

Hence, the objective of the defender is formulated as minimizing the expected cumulative cost $C_t$ over the time horizon $T$ with discount factor $\gamma = 1$: 

\begin{equation}
\label{eq:pomdp-obj-func}
    J = \sum_{t=1}^T \mathds{E} [C_t]
\end{equation}

\subsubsection{The defender's optimization problem}

We define $\pi_D: \mathcal{H} \to \textset{D}$ or $\pi_D:\textset{B} \to \textset{D}$ as the defender strategy, where $\mathcal{H}$ is the history space and $\textset{B}$ is the belief space. The defender problem is defined as finding the optimal strategy $\pi_D^*$ that minimizes the expected cumulative cost $J$.

\begin{problem}
\label{prob:pomdp}
Find the optimal defender strategy :
\begin{subequations}
\label{eq:pomdp-opt-problem}
\begin{align}
    \pi_D^* \; \; \;\;\;& = \; \argmin_{\pi_D} \mathds{E}_{\pi_D} [J] \\
    \text{subject to } \; \; \;\;\; & D_t = \pi_D(b_t)  \forall t \\
                       & \pi_A \sim P(\Pi_A)
\end{align}
\end{subequations}
\end{problem}

Because the POMDP has finite state, action, and observation spaces over a finite horizon, an optimal strategy $\pi^*_D$ exists~\cite[Thm. 7.4.1]{pomdp}.

%% --------------- Learning methods ----------------------------
%% ----------------------------------------------------------------

\section{Learning Defender Strategies with Reinforcement Learning}
\label{sec:methods}

Problem~\ref{prob:pomdp} can be solved with Dynamic Programming methods~\cite{dynamic-programming-pomdp}. Due to the large state space, such methods are computationally intractable for our use case ~\cite{pomdp-complexity2}. We therefore parameterize the defender strategy $\pi_D$ with parameter $\theta$. We apply PPO~\cite{ppo}, a state-of-the-art RL algorithm, to learn the (almost) optimal defender strategy. PPO is an actor--critic policy-gradient method that uses a clipped surrogate objective to improve training stability by limiting policy changes between updates. 

% To find the optimal policy in a POMDP, two main classes of methods can be used: dynamic programming and Reinforcement Learning. Dynamic programming methods~\cite{dynamic-programming-pomdp} have a strong theoretical foundation and in principle can compute optimal strategies. However, due to the large size of the state space of our POMDP model, dynamic programming methods are computationally intractable~\cite{pomdp-complexity1, pomdp-complexity2}.

% We therefore use the second class of methods, reinforcement learning (RL), to learn optimal defender strategies. Specifically, we use Proximal Policy Optimization (PPO)~\cite{ppo}, a state-of-the-art RL algorithm, to learn a parameterized policy $\pi_\theta$. The policy maps an input representation $x_t$ to a probability distribution over defender actions: $\pi_\theta(x_t) \in \Delta(D)$. PPO is a policy-gradient method with an actor--critic architecture. It learns the policy by applying gradient descent with the following gradient:

% \begin{equation}
%     \nabla_\theta J(\theta)=\mathbb{E}_{\pi_\theta}\left[\nabla_\theta \log\pi_\theta(D_t\mid x_t) , A^{\pi_\theta}(x_t,D_t) \right],
% \end{equation}

% \noindent where $A^{\pi_\theta}(x_t,D_t)$ is the advantage function. In PPO, this update is performed with a clipped objective that limits the change between consecutive policies, improving training stability.

A direct application of PPO to Problem~\ref{prob:pomdp} is not feasible due to the large state space and belief space. We therefore propose two PPO-based solution methods that use different techniques to address the issue.

% In this paper, we present two solution methods to learn defender strategies. They are based on PPO and differ in the input to the neural network that represents the defender strategy. 

\subsection{PPO with $k$ latest observations: $k$-Obs-PPO}

The first method, $k$-Obs-PPO, uses the $k$ most recent IDS observations, $(O_t, \dots,O_{t-k+1})$ as input for $\pi^\theta_D$. It allows training the defender strategy directly from observation histories, without computing the belief state. The parameter $k$ controls the trade-off between the amount of information used to learn the strategy on the one hand and the computational complexity and sampling efficiency on the other hand. A larger $k$ provides more temporal information to the defender strategy but increases the input dimension and the learning complexity. A smaller $k$ keeps the input small and simplifies defender strategy learning. 

% The first method, which we call $k$-Obs-PPO, uses the $k$ most recent IDS observations, $(O_t, \dots,O_{t-k+1})$ as input for $\pi^\theta_D$. The method is simple and allows training the strategy directly from observation histories without computing or updating the belief state. The parameter $k$ controls the trade-off between the information used to learn the strategy on one hand, and the computational complexity and sampling efficiency on the other hand. A larger $k$ provides more temporal information to the defender strategy but increases input dimension and learning complexity. A smaller $k$ keeps the input small and simplifies policy learning. 

\subsection{Belief Filter Proximal Policy Optimization: BF-PPO}

The second method, BF-PPO, uses an approximate and simplified belief state to learn the defender strategy. In a POMDP, belief updates are computed using Bayes filter (Eq.~\eqref{eq:belief-update}). Such an update has a quadratic time complexity with respect to the size of the state space. Therefore, the Bayes Filter is computationally infeasible in our case. We address the issue by approximating the belief using particle filter~\cite{particle_filter}. At time $t$, the particle filter represents the belief by $M$ sampled states, $\mathcal{P}_t = \{s_t^{(1)},\dots,s_t^{(M)}\}$. The belief state is then approximated by the relative frequency of each system state in $\mathcal{P}_t$, i.e., $\hat{b}_t(s) = \frac{1}{M}\sum_{i=1}^{M}\mathds{1}\{s_t^{(i)} = s\}$.

Due to the large state space, the belief vector $\langle \hat b_t(s)\rangle$ is high dimensional, which increases the learning complexity and makes it inefficient as input to the defender strategy. We therefore exploit the factorized structure of the system state with components $\hoststate{0}{t},\hoststate{1}{t}, \hoststate{2}{t}, \procstate{0}{t},\procstate{1}{t},A_{t-1}$. For each such component $X_t$ of the state and component state $x$, we compute the marginal belief
\begin{equation}
    \hat{b}_t^X(x)=\frac{1}{M}\sum_{i=1}^{M}\mathds{1}\{X_t^{(i)} = x\}.
\end{equation}

The concatenation of these marginal beliefs is used as input to the defender strategy $\pi^\theta_D$, providing a compact input while preserving the information about each component of the system state.

%% --------------- Emulation ----------------------------
%% ----------------------------------------------------------------

\section{Constructing an Emulation Environment and a Simulation Environment to Train and Evaluate Defender Strategies}
\label{sec:emulation}

Learning defender strategies using our solution methods and evaluating them requires repeated POMDP episodes. Such episodes cannot be performed directly in an operating OT infrastructure, which is both inefficient due to the large number of episodes required for training and impractical because it requires performing attacker actions in the system. For these reasons, we develop an emulation environment and a simulation environment. The emulation environment is designed to closely match the operational conditions of the target infrastructure. It serves two purposes: collecting IDS measurements to identify the observation model of the POMDP, and evaluating defender strategies under realistic conditions. The simulation environment generates POMDP episodes to efficiently train the defender strategies.

\subsection{Constructing the Emulation Environment}

The infrastructure in Fig. \ref{fig:ot-topo} is emulated with Docker~\cite{docker} containers and orchestrated by ContainerLab~\cite{containerlab}. Each functional component is instantiated as a separate container, including the enterprise clients, supervisory hosts, PLCs, water tanks, router, attacker, defender, and IDS. The PLCs are implemented using OpenPLCv3~\cite{openplc}. The main communication protocol between OT components (HMI, PLC, EWS, and water tanks) is implemented using ModbusTCP~\cite{modbus}. The configuration of the components is listed in Appendix~\ref{sec:appendix-emulation}.

% Each functional component in the OT system is instantiated as a separate container, including the enterprise clients, supervisory hosts, PLCs, water tanks, router, attacker, defender and IDS. The PLCs at the control subnet are implemented using OpenPLCv3~\cite{openplc} with a control program for interacting with the corresponding water tanks. The configuration of the containers are presented in Tab. \ref{tab:container-specs}.

The network is implemented using Linux virtual Ethernet interfaces. Logical network segmentation is enforced through VLAN configurations using Open vSwitch~\cite{ovs-switch}. Inter-subnet routing and firewall rules are implemented by the router container. The IDS is connected to a mirrored Open vSwitch port, which allows it to observe and collect network measurements.

% The network is emulated using Linux virtual Ethernet interfaces, while logical subnet separation is implemented through VLANs configuration with Open vSwitch~\cite{ovs-switch}. Inter-subnet communication and basic firewall rules are implemented by a router container. The IDS is connected to a mirrored Open vSwitch port, allowing it to observe inter-subnet traffic and collect the network measurements used to construct the defender observation. Communication between the supervisory, control, and physical layers is based on ModbusTCP~\cite{modbus}. HMIs communicate with PLCs for monitoring and control, and PLCs exchange sensor and actuator values with the water-tank processes through Modbus registers and coils. The EWS is also connected to the PLCs, mainly through HTTP protocol. T

The defender's reset action on a host or a process changes the IP address and login credentials of the target component, thereby removing potential attacker access. The updated configuration is then propagated to the components that require it. For example, when a PLC is reset, its new IP address is distributed to the HMIs and the EWS so that supervisory communication can be re-established.

% The defender can reset either a supervisory host or a physical process. Each reset triggers a reconfiguration procedure that changes the IP address and login credentials of the retested component and propagates the updated configuration to the relevant components. For example, when a PLC is reseted, its new IP address is distributed to the HMIs and the EWS so that supervisory communication can be re-established. 

The attacker actions, including reconnaissance and exploitation, are implemented by a sequence of commands executed from the attacker container. To capture a broad range of traffic patterns for the same action, the implementation varies command parameters, including target networks and ports, probe types, command arguments, and payload sizes. This allows us to observe traffic under a wide range of attacker behaviors. More details of the implementation are presented in Appendix~\ref{sec:appendix-emulation}.

% At each time step, the attacker executes one action sampled from the attacker action space, as summarized in Tab.~\ref{tab:attacker-action}. Each abstract action is mapped to a concrete command sequence in the emulation, including network scanning, host scanning, HMI and EWS exploitation, host inspection, and process tampering. The implementation introduces variability through stealth profiles and action-specific parameters, such as target networks and ports, probe types, command arguments, and payload sizes. This yields multiple traffic realizations for the same abstract attacker action, enabling estimation of observation distributions under different attacker behaviors.

Background traffic is generated by three types of activities: by enterprise clients using services on the HMI and EWS containers; by HMI-PLC communication, where HMIs periodically read process measurements from the PLCs and issue routine control commands; and by EWS-PLC communication, representing engineering operations performed by an operator. All non-periodic background activities are generated by independent Poisson processes.

% Background traffic is generated by three classes of benign activities. The first class consists of enterprise clients accessing services on the HMI and EWS containers to perform normal functional operations. The second class consists of HMI--PLC communication, where HMIs periodically read process measurements from the PLCs and issue routine control commands, including local operator actions at the HMI. The third class consists of local EWS--PLC communication, representing engineering operations performed by an operator. All non-periodic background activities are generated by independent Poisson processes.

\subsection{Estimating the observation function in the emulation environment}

In the emulation environment, we perform operation in periods of $30$ seconds wall clock time. During such a period, the attacker performs an attack action according to its strategy. Also, the defender performs a defensive action based on the observation data collected during the period. This data consists of network statistics obtained from the IDS, namely, $o = (\vgobs{}, \vlobshost{0}{}, \vlobshost{1}{}, \vlobshost{2}{} ,\vlobsproc{0}{}, \vlobsproc{1}{})$ (Eq.~\eqref{eq:pomdp-observation}).

% In the emulation environment, the monitoring period is $30$ seconds. During such a period, the attacker performs an attack action according to its strategy. The defender performs a defensive action based on the observation collected during the period. This observation consists of network statistics obtained from the IDS, specifically, $o = (\vgobs{}, \vlobshost{0}{}, \vlobshost{1}{}, \vlobshost{2}{} ,\vlobsproc{0}{}, \vlobsproc{1}{})$ (Eq.~\eqref{eq:pomdp-observation})

\begin{figure}[h!]
    \centering
    \begin{subfigure}[b]{0.45\textwidth}
      \centering
      \includegraphics[width=0.95\linewidth]{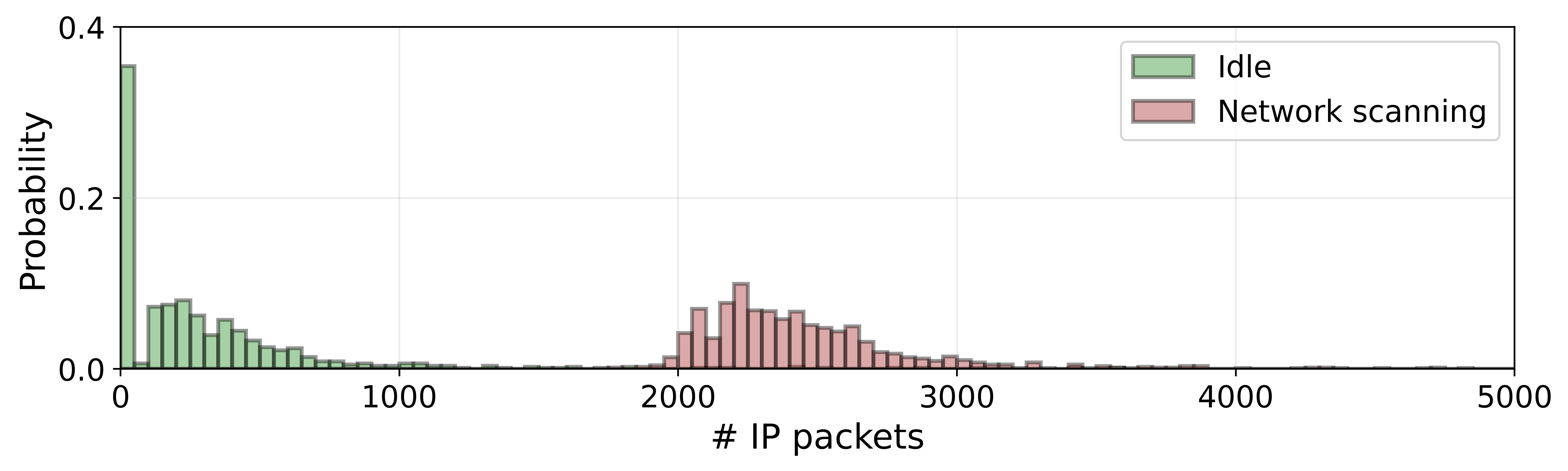}
    \end{subfigure}
    \begin{subfigure}[b]{0.45\textwidth}
      \centering
      \includegraphics[width=0.95\linewidth]{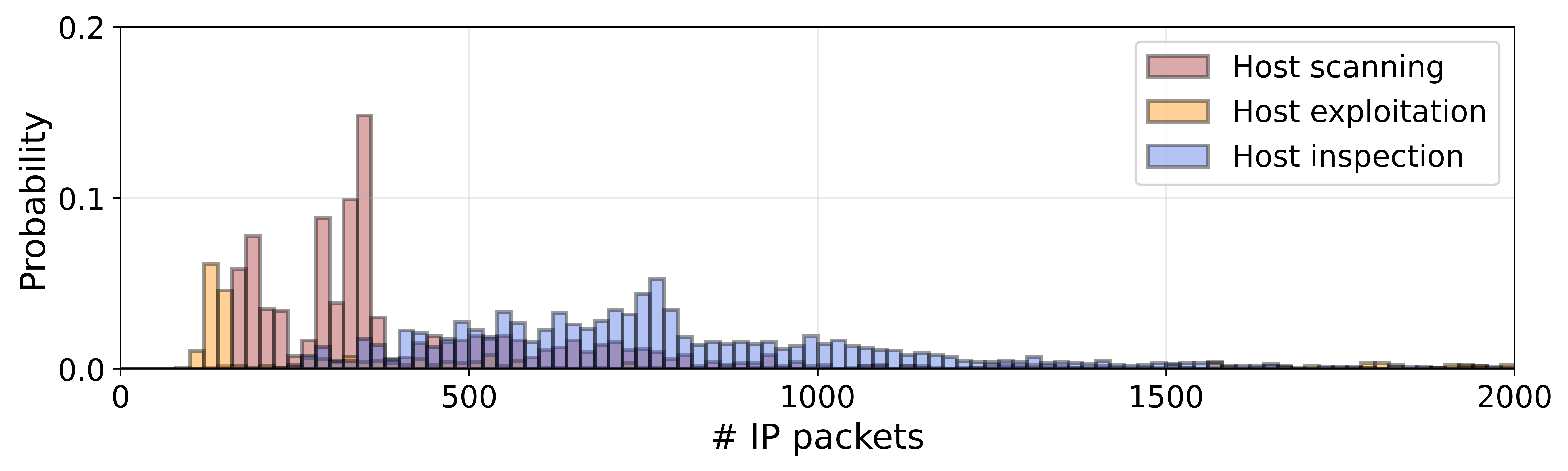}
    \end{subfigure}
    \begin{subfigure}[b]{0.45\textwidth}
      \centering
      \includegraphics[width=0.95\linewidth]{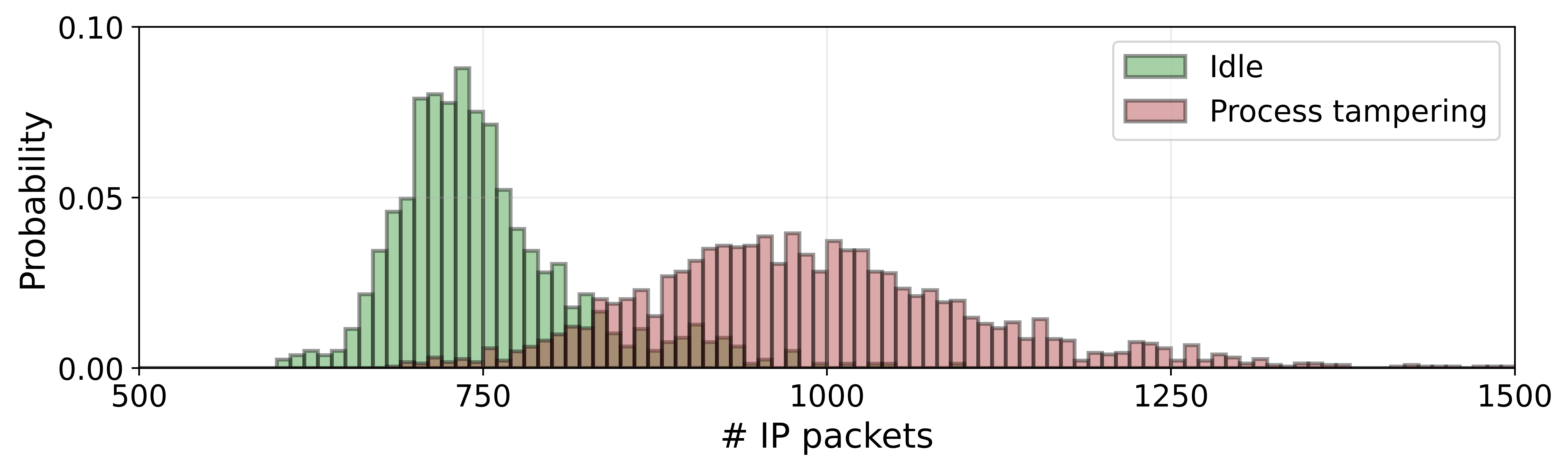}
    \end{subfigure}
    \caption{Empirical distribution of IP packet counts measured on the switch in Fig.~\ref{fig:ot-topo}. The top figure relates to  the total traffic passed through the switch; the middle figure relates to the traffic between EWS-0 and entities in the enterprise subnet; the lower figure relates to the traffic between PLC-0 and entities in the supervisory subnet. Different colors refer to measurements from different attacker actions.}
    \label{fig:eg-distribution}
\end{figure}

To estimate the observation function $\textfunc{Z}$ (Eq.~\eqref{eq:pomdp-observation-function}) of the POMDP, we collect monitoring observation data from $40000$ periods, which amounts to $14$ days. Effectively estimating $Z(o,s,d)=\Pr(O_{t+1}=o \mid S_{t+1}=s,D_t=d)$ is not feasible since it would require approximately $10^8$ observations ($|\mathcal{S}||\mathcal{D}| = 52500$ possible conditioning combinations of $(s,d)$ and assuming about $2000$ observations needed for each combination). To work with $40000$ observation measurements, we simplify the observation function $\textfunc{Z}$ to $\hat{\textfunc{Z}}$ by setting $\hat{\textfunc{Z}}(o,a) = \Pr(O_{t+1}=o \mid A_{t}=a)$, where $a$ denotes the attacker action. Fig.~\ref{fig:eg-distribution} shows empirical distributions for selected attacker actions.

% To estimate the observation function $\textfunc{Z}$ (Eq.~\eqref{eq:pomdp-observation-function}) of the POMDP, we collect monitoring observation data from $40000$ periods, which amounts to 14 days. Due to the large state space, directly estimating $Z(o,s,d)=\Pr(O_{t+1}=o \mid S_{t+1}=s,D_t=d)$ is sampling inefficient. Therefore, we estimate $\textfunc{Z}(o,s,d)$ with $\hat{\textfunc{Z}}(o,a) = \Pr(O_{t+1}=o \mid A_{t}=a)$, where $a$ denotes the abstract attacker action, including \textit{Idle} ($A^h_t=\emptyset, A^p_t=\emptyset$),  \textit{Network scanning} ($A^h_t=\textset{H}$), \textit{Host scanning} ($A^h_t=\textset{h}, S^h_t=\textstate{D}$), \textit{Host exploitation} ($A^h_t=\textset{h}, S^h_t=\textstate{S}$), \textit{Host inspection} ($A^h_t=\textset{h}, S^h_t=\textstate{E}$), and \textit{Process tampering} ($A^h_t=\textset{h}, A^p_t=\textset{p}, S^h_t=\textstate{I}$). 

\subsection{Running the POMDP simulation}

Simulating a POMDP episode starts the system in state $s = (\textstate{U}, \textstate{U}, \textstate{U}, \textstate{W}, \textstate{W}, (\emptyset, \emptyset))$. At each time step, the attacker and the defender perform an action following their respective strategies. The system state is then updated following the transition function $\textfunc{T}$ specified in~\secref{sec:pomdp}. The observation that the defender makes during the time step is sampled from $\hat{\textfunc{Z}}$. Further, the cost incurred for this time step is computed using the cost function $\textfunc{C}$. The episode ends when the $T$ time steps have been executed. The trajectories of observations, actions, and costs per period are used to update the defender strategies during training.

% Each episode in the POMDP simulation environment starts with the initial distribution $\rho_1$. At each time $t$, the attacker take action $A_t$ and defender takes action $D_t$ following their respective strategies. Based on the actions, the state of the system is updated from $S_t$ to $S_{t+1}$ according to the transition function $\textfunc{T}$ specified in~\secref{sec:pomdp}. The simulation environment computes the cost $C_t$ according to the cost function $\textfunc{C}$. Finally, the defender observation $O_{t+1}$ at time $t+1$ is sampled from $\hat{\textfunc{Z}}$. The episode proceeds until the finite horizon $T$ is reached. The generated trajectories of observations, actions, and costs are used to update the defender strategies during training.

%% --------------- Evaluation ----------------------------
%% ----------------------------------------------------------------

\section{Computing the Defender Strategies in the Simulation Environment and Evaluating Them in the Emulation Environment}
\label{sec:evaluation}

We train BF-PPO and two $k$-Obs-PPO variants, with $k = 1$ and $k=4$ in the simulation environment. We then evaluate the obtained defender strategies in the emulation environment. We compare the proposed methods with two baselines, a threshold-based defender strategy and an idealized PPO baseline, where we assume the defender has full observation of the system states and the attacker actions. The training and evaluation are conducted on an Apple M3 Pro processor. The hyperparameters for the learning methods are listed in Appendix~\ref{sec:appendix-hyperparameters}. The source code is available at~\cite{implement-github}.

%%%%%%%%%%%%%%%%%%%%%%%%%%%%%% Setup %%%%%%%%%%%%%%%%%%%%%%%%%%%%%%%%%%%%
%%%%%%%%%%%%%%%%%%%%%%%%%%%%%% Setup %%%%%%%%%%%%%%%%%%%%%%%%%%%%%%%%%%%%
%%%%%%%%%%%%%%%%%%%%%%%%%%%%%% Setup %%%%%%%%%%%%%%%%%%%%%%%%%%%%%%%%%%%%

\subsection{Evaluation setup}

\subsubsection{Attacker strategies}

We train and evaluate the methods against three attacker strategies:

\paragraph{Opportunistic} The attacker prioritizes reaching process tampering as quickly as possible. It selects actions that advance the attack along the shortest available path toward tampering with a process.

\paragraph{Explorative} The attacker prioritizes gathering knowledge about the infrastructure before tampering with a process. It scans supervisory hosts, exploits available entry points, and inspects host--process control relationships before tampering with the processes.

\paragraph{Adaptive} This attacker adapts to defender actions by avoiding recently reset components and switching to alternative attack paths when possible.

\subsubsection{Baselines}

\paragraph{MDP-PPO} The method applies PPO whereby the system state can be observed by the defender. This baseline method represents an idealized setting and a cost lower bound for the PPO-based methods proposed in this paper.

\paragraph{Threshold strategy}

At each time step $t$, the defender computes a cost score $R^t_h$ for each host and a cost score $R^t_p$ for each process. If the score of any process exceeds its respective threshold, the defender following this strategy resets the process with the highest score. Otherwise, if the score of any host exceeds its respective threshold, the defender resets the host with the highest score. The thresholds are determined by the reset costs $\sigma_d$ (see Tab.~\ref{tab:cost-parameter}). The cost scores are given by the formulas in Appendix~\ref{sec:appendix-threshold-baseline}.

% A defender following this strategy resets the process with the highest score, if any process exceeds the its respective threshold. Otherwise, the defender resets the host with the highest score, if any host exceeds the threshold. The thresholds are defined by the reset costs $\sigma_d$ (see Tab.~\ref{tab:cost-parameter}). The cost scores are given in Appendix~\ref{sec:appendix-threshold-baseline}.

\subsubsection{Evaluation process} We train the defender strategies using 
BF-PPO, $1$-Obs-PPO, and $4$-Obs-PPO, as well as the baseline method MDP-PPO in the simulation environment. For each method, we perform four training runs with different seeds. A training run consists of $500$ iterations, where each iteration contains $100$ episodes with time horizon $100$. After each iteration, we evaluate the current policy to obtain a point in the learning curve.

We finally evaluate the learned defender strategies in the emulation environment. The strategy obtained by each method is evaluated against the three attacker strategies for $20$ episodes with time horizon $100$. Evaluating a defender strategy against an attacker strategy takes about $17$ hours on an Apple M3 Pro processor.

%%%%%%%%%%%%%%%%%%%%%%%%%%%%%% Result %%%%%%%%%%%%%%%%%%%%%%%%%%%%%%%%%%%%
%%%%%%%%%%%%%%%%%%%%%%%%%%%%%% Result %%%%%%%%%%%%%%%%%%%%%%%%%%%%%%%%%%%%
%%%%%%%%%%%%%%%%%%%%%%%%%%%%%% Result %%%%%%%%%%%%%%%%%%%%%%%%%%%%%%%%%%%%

\bgroup
\begin{figure*}[htbp]
    \centering
    \includegraphics[width=0.97\linewidth]{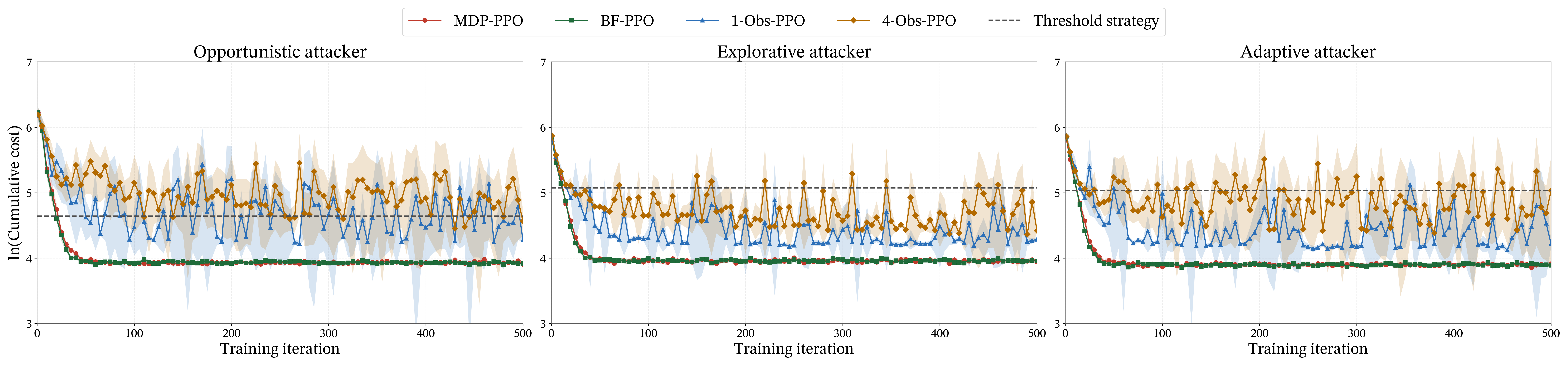}
    \caption{The learning curves for the solution methods proposed in~\secref{sec:methods} (BF-PPO -- green curves, $1$-Obs-PPO -- blue curves, and $4$-Obs-PPO -- orange curves) and the baselines (MDP-PPO -- red curves and Threshold strategy -- dashed lines). The solution methods computed effective defender strategies for three different attacker strategies (Opportunistic, Explorative and Adaptive). The learning curves show the average and the confidence intervals of $120$ episodes. The best performing solution method with respect to convergence speed and cost is BF-PPO.}
    \label{fig:learning-curves}
\end{figure*}
\egroup
\bgroup
\begin{figure*}[htbp]
    \centering
    \includegraphics[width=0.97\linewidth]{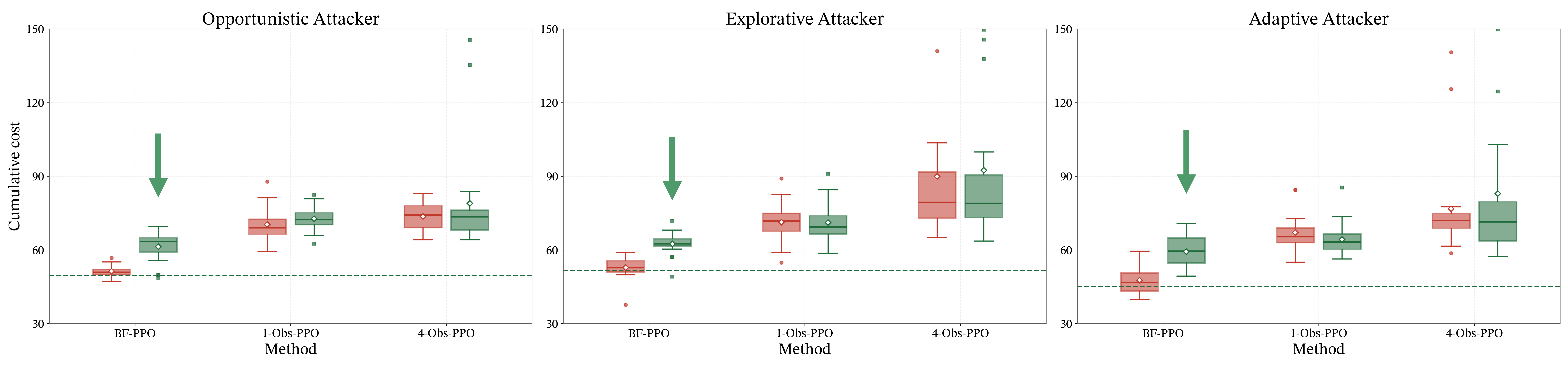}
    \caption{Evaluation of the learned defender strategies for the proposed solution methods, BF-PPO, $1$-Obs-PPO and $4$-Obs-PPO. The green box plots relate to evaluation results in the emulation environment. For comparison, we add the results from the simulation environment, presented by red box plots. The baseline refers to the learned defender strategies under the assumption of full observability of the system states.}
    \label{fig:evaluation}
\end{figure*}
\egroup

\subsection{Evaluating the methods in the simulation environment}

Fig.~\ref{fig:learning-curves} shows the learning curves of the proposed methods and the baselines against three attacker strategies. BF-PPO, $1$-Obs-PPO, and $4$-Obs-PPO, are represented by the green, blue, and orange curves, respectively. MDP-PPO and Threshold strategy are shown by red curves and dashed lines, respectively. Each plot corresponds to a attacker strategy and shows the natural logarithm of the cumulative cost during training.

The learning curves show that $1$-Obs-PPO and $4$-Obs-PPO converge to higher cumulative costs than BF-PPO for all attacker strategies. They also exhibit larger variation. $1$-Obs-PPO tends to perform better than $4$-Obs-PPO, which suggests that, in this setting, increasing the observation history does not provide sufficient additional information to improve performance. 

BF-PPO converges faster and to lower cumulative costs than the $k$-Obs-PPO variants for all attacker strategies. Its learning curves are also close to those of MDP-PPO, the full-observability baseline. This indicates that using particle filter and marginal beliefs provides a solution for learning effective defender strategies. Further, Fig.~\ref{fig:learning-curves} also shows that BF-PPO consistently outperforms the Threshold strategy. 

\subsection{Evaluating the defender strategies in the emulation environment}

Fig.~\ref{fig:evaluation} shows the evaluation results of the defender strategies in the emulation environment (green color) and compares them with the results from the simulation environment (red color). The box plots show the distributions of cumulative costs over $20$ evaluation episodes for BF-PPO, $1$-Obs-PPO and $4$-Obs-PPO. The dashed line shows the average cumulative cost of MDP-PPO, the full-observability baseline method.

The plots show that BF-PPO outperforms the $k$-Obs-PPO variants for all attacker strategies. Its cumulative costs are also close to those of MDP-PPO, which is consistent with the simulation results in Fig.~\ref{fig:learning-curves}.

%The k-Obs-PPO variants show similar average performance in simulation and emulation, suggesting that their learned strategies transfer consistently between the two environments. However, their performance is less stable in the emulation environment, especially for $k=4$. 

BF-PPO shows a gap between the performance in the emulation environment and the simulation environment. We expect that the defender strategy performs less well in the emulation environment than in the simulation environment. This is due to the fact that the formal model we have presented is an approximation of the more realistic emulated system, and the defender strategy has been optimized for the simulation environment.

Overall, we find that BF-PPO is the strongest proposed method for our use case and the considered attacker strategies.

%% --------------- Conclusions ----------------------------
%% ----------------------------------------------------------------

\section{Conclusion and Future Work}
\label{sec:conclusion}

Based on the POMDP model of the use case, we have defined an optimal defender strategy and proposed two solution methods for learning the approximation of this strategy. The solution methods allow us to learn the defender strategies at a reasonable cost and with low sample complexity. The evaluation in the emulation environment shows that effective defender strategies can be learned with these methods. In particular, BF-PPO approaches the performance of a baseline with full observability.

As for future work, we plan to evaluate the proposed model and solution methods on an industrial testbed in collaboration with an industrial partner, for an intrusion response use case similar to the one studied in this paper. This evaluation will allow us to study our approach to modeling and computing defender strategies in a different and realistic environment and further to include safety constraints.

%% --------------- Acknowledgement ----------------------------
%% ----------------------------------------------------------------

\section{Acknowledgment}

This work has been supported by the DARPA CASTLE program through project ORLANDO and by the WASP NEST program through project AIRR. The authors thank KTH researchers Kim Hammar and Xiaoxuan Wang for their constructive comments.

%\label{sec:acknoledgement}
%\input{parts/acknowledgement}

%% ---------------  Appendix ----------------------------
%% ----------------------------------------------------------------
% \section*{Appendix}
% \label{sec:appendix}

\section*{Appendix}
\addcontentsline{toc}{section}{Appendix}

\setcounter{subsection}{0}
\renewcommand{\thesubsection}{\Alph{subsection}}
\subsection{Emulation configurations: Tab. \ref{tab:container-specs} and Tab. \ref{tab:attacker-action}}
\label{sec:appendix-emulation}

\subsection{Training hyperparameters: Tab. \ref{tab:appendix-hyperparam}}
\label{sec:appendix-hyperparameters}

\bgroup
\def\arraystretch{1.1} %Change margin inside cells
\begin{table}[ht]
    \centering 
    \caption{Configuration of the emulated physical components}
    \tiny
    \begin{tabular}{ccp{0.2\textwidth}}
    \toprule
        \textbf{Component} & \textbf{Services} & \textbf{Vulnerabilities} \\
        \midrule
        % $\mathrm{Client}_0, \mathrm{Client}_1, \mathrm{Attacker}$ & -- & -- \\
        % \midrule
        $\mathrm{HMI}_0, \mathrm{HMI}_1$  & HTTP & Weak credentials  \\
        \midrule
        $\mathrm{EWS}_0$ &  SSH, Telnet, SMB & SSH/Telnet weak credentials, CVE-2017-7494 \\
        \midrule
        $\mathrm{PLC}_0, \mathrm{PLC}_1$  & ModbusTCP, HTTP & -- \\
        \midrule
        $\mathrm{Tank}_0, \mathrm{Tank}_1$ & ModbusTCP & -- \\
    \bottomrule
    \end{tabular}
    \label{tab:container-specs} 
\end{table}
\egroup

\bgroup
\def\arraystretch{1.1}
\begin{table}[ht]
\tiny
\centering
\caption{Implementation of attacker actions in the emulation}
\begin{tabular}{cp{0.3\textwidth}}
\toprule
\textbf{Attacker action} & \textbf{Implementation} \\
\midrule
Scan network & ICMP scan \\
\midrule
Scan host & TCP/UDP/OS scan, lightweight probes \\
\midrule
Exploit host & HMI: HTTP brute-force, SQL injection, EWS: SSH/Telnet brute-force, SambaCry\\
\midrule
Inspect host & HMI: HTTP probes, EWS: ModbusTCP control \\
\midrule
Tamper process & HMI: PLC parameter modifications via web interface, EWS: Craft \& Upload PLC programs, PLC parameter control via ModbusTCP\\
\bottomrule
\end{tabular}
\label{tab:attacker-action}
\end{table}
\egroup

\bgroup
\def\arraystretch{1.1} %Change margin inside cells
\begin{table}[ht]
    \centering
    \scriptsize
    \caption{Hyperparameters for the training of defender strategies}
    \begin{tabular}{p{0.22\textwidth}p{0.2\textwidth}}
    \toprule
        \textbf{Parameter} & \textbf{Value} \\ 
    \midrule
    \# particles (BF-PPO) $M$ & $500$ \\
    \# hidden layer, \# neurons, random seeds & $2, 128, [77, 108, 433, 841]$ \\
    l.r. $\alpha$, $\gamma$, GAE $\lambda$, ENT-COEF, clip $\epsilon$  & $0.0001, 0.99, 0.95, 0.001, 0.2$ \\
    Optimiser (parameters)  & Adam ($\beta_1 = 0.9, \beta_2 = 0.99)$\\
    
    \bottomrule

    \end{tabular}
    \label{tab:appendix-hyperparam} 
\end{table}
\egroup

% Tab.~\ref{tab:container-specs} presents the configurations of the hosts in the emulation environment. Tab.~\ref{tab:attacker-action} presents the emulated attacker actions in the emulation environment. Tab.~\ref{tab:appendix-hyperparam} presents the hyper-parameters used to learn defender strategies.

\subsection{Threshold strategy's cost score formulas}
\label{sec:appendix-threshold-baseline}
Eq.~\eqref{eq:threshold-baseline} defines the formulation of the cost score for each host and process at time $t$, where $b_t$ denotes the belief state.
\begin{equation}
\label{eq:threshold-baseline}
R^t_h = \sum_{s\in S^h} b_t^h(s)\sigma_h(s); \quad R^t_p = \sum_{s\in S^p} b_t^p(s)\sigma_p(s)
\end{equation}

%% --------------- References -----------------------------------
%% -----------------------------------------------------------------

% \printbibliography

\bibliographystyle{IEEEtran}
\bibliography{refs}

@book{particle_filter,
  title={Probabilistic Robotics},
  author={Thrun, S. and Burgard, W. and Fox, D.},
  isbn={9780262201629},
  lccn={2005043346},
  year={2005},
  publisher={MIT Press}
}

@misc{ppo,
      title={Proximal Policy Optimization Algorithms}, 
      author={John Schulman and Filip Wolski and Prafulla Dhariwal and Alec Radford and Oleg Klimov},
      year={2017},
      eprint={1707.06347},
      archivePrefix={arXiv},
      primaryClass={cs.LG},
}

@book{pomdp, 
place={Cambridge}, 
title={Partially Observed Markov Decision Processes: From Filtering to Controlled Sensing}, 
publisher={Cambridge University Press}, 
author={Krishnamurthy, Vikram}, 
year={2016}}

@article{pomdp-complexity2,
title = {On the complexity of partially observed Markov decision processes},
journal = {Theoretical Computer Science},
volume = {157},
number = {2},
pages = {161-183},
year = {1996},
issn = {0304-3975},
doi = {https://doi.org/10.1016/0304-3975(95)00158-1},
author = {Dima Burago and Michel {de Rougemont} and Anatol Slissenko},
}

@article{dynamic-programming-pomdp,
 ISSN = {0030364X, 15265463},
 author = {Edward J. Sondik},
 journal = {Operations Research},
 number = {2},
 pages = {282--304},
 publisher = {INFORMS},
 title = {The Optimal Control of Partially Observable Markov Processes Over the Infinite Horizon: Discounted Costs},
 volume = {26},
 year = {1978}
}

@book{ics-security-book,
  title={Industrial Cybersecurity: Efficiently monitor the cybersecurity posture of your ICS environment},
  author={Ackerman, Pascal},
  year={2021},
  publisher={Packt Publishing Ltd}
}

@misc{nist-ot-security,
  title     = {Guide to Operational Technology (OT) Security},
  author    = {Stouffer, Keith and Pease, Michael and Tang, CheeYee and Zimmerman, Timothy and Pillitteri, Victoria and Lightman, Suzanne and Hahn, Adam and Saravia, Samuel and Sherule, Ajit and Thompson, Michael},
  year      = {2023},
  month     = {9},
  publisher = {National Institute of Standards and Technology},
  address   = {Gaithersburg, MD},
  type      = {NIST Special Publication (SP)},
  number    = {NIST SP 800-82r3},
  doi       = {10.6028/NIST.SP.800-82r3},
}

@misc{isa-iec-guideline,
  author       = {{International Society of Automation}},
  title        = {{IEC 62443 Series of Standards}},
  year         = {2026},
  howpublished = {\url{https://www.isa.org/standards-and-publications/isa-standards/isa-iec-62443-series-of-standards}},
}

@article{purdue-model,
title = {The Purdue enterprise reference architecture},
journal = {Computers in Industry},
volume = {24},
number = {2},
pages = {141-158},
year = {1994},
issn = {0166-3615},
doi = {https://doi.org/10.1016/0166-3615(94)90017-5},
author = {Theodore J. Williams},
}

@article{mitre,
author = {Al-Sada, Bader and Sadighian, Alireza and Oligeri, Gabriele},
title = {MITRE ATT\&CK: State of the Art and Way Forward},
year = {2024},
issue_date = {January 2025},
publisher = {Association for Computing Machinery},
address = {New York, NY, USA},
volume = {57},
number = {1},
issn = {0360-0300},
doi = {10.1145/3687300},
journal = {ACM Comput. Surv.},
month = oct,
articleno = {12},
numpages = {37}
}

@article{it-ot-convergence1,
  title={Systematic review and characterisation of malicious industrial network traffic datasets: M. Dobler et al.},
  author={Dobler, Martin and Hellwig, Michael and Lopes, Nuno and Oakley, Ken and Winterburn, Mike},
  journal={International Journal of Information Security},
  volume={24},
  number={5},
  pages={208},
  year={2025},
  publisher={Springer}
}

@article{it-ot-convergence2,
   title={Industrial and Critical Infrastructure Security: Technical Analysis of Real-Life Security Incidents},
   volume={9},
   ISSN={2169-3536},
   DOI={10.1109/access.2021.3133348},
   journal={IEEE Access},
   publisher={Institute of Electrical and Electronics Engineers (IEEE)},
   author={Makrakis, Georgios Michail and Kolias, Constantinos and Kambourakis, Georgios and Rieger, Craig and Benjamin, Jacob},
   year={2021},
   pages={165295–165325}}

@techreport{waterfall-report,
  title       = {2024 Threat Report: OT Cyberattacks with Physical Consequences},
  author      = {{Waterfall Security Solutions} and {ICS Strive}},
  year        = {2024},
  institution = {Waterfall Security Solutions},
  type        = {Technical Report},
}

@techreport{dragos-report,
  title       = {2025 OT/ICS Cybersecurity Report: A Year in Review},
  author      = {{Dragos, Inc.}},
  year        = {2025},
  institution = {Dragos Inc.},
  type        = {Technical Report},
}

@article{incident-triton,
  title={TRITON: The first ICS cyber attack on safety instrument systems},
  author={Di Pinto, Alessandro and Dragoni, Younes and Carcano, Andrea},
  journal={Proc. Black Hat USA},
  volume={2018},
  pages={1--26},
  year={2018}
}

@techreport{incident-industroyer,
  title={PRECURSOR ANALYSIS REPORT: INDUSTROYER2 AND WIPER MALWARE TARGETING UKRAINIAN ENERGY PROVIDER 2022},
  author={McCarthy, Daniel Dennis},
  year={2025},
  institution={Idaho National Laboratory (INL), Idaho Falls, ID (United States)}
}

@article{ot-cybersec-weakness,
  title={Analysis of Publicly Accessible Operational Technology and Associated Risks},
  author={Rodda, Matthew and Mavroudis, Vasilios},
  journal={arXiv preprint arXiv:2508.02375},
  year={2025}
}

@INPROCEEDINGS{related-work-rule-based,
  author={Vaudey, Jolahn and Mocanu, St{\'e}phane and Delaval, Gwena{\"e}l and Rutten, Eric},
  booktitle={2025 IEEE Conference on Communications and Network Security (CNS)}, 
  title={Reconfiguration of Firewall Filter Rules as a Response to Industrial Control System Intrusion}, 
  year={2025},
  volume={},
  number={},
  pages={1-6},
  doi={10.1109/CNS66487.2025.11195004}
}

@ARTICLE{related-work-rule-based2,
  author={Murillo Piedrahita, Andr{\'e}s F. and Gaur, Vikram and Giraldo, Jairo and C{\'a}rdenas, {\'A}lvaro A. and Rueda, Sandra Julieta},
  journal={IEEE Software}, 
  title={Leveraging Software-Defined Networking for Incident Response in Industrial Control Systems}, 
  year={2018},
  volume={35},
  number={1},
  pages={44-50},
  doi={10.1109/MS.2017.4541054}
}

@article{related-work-rule-based3,
  title   = {Virtual incident response functions in control systems},
  journal = {Computer Networks},
  volume  = {135},
  pages   = {147--159},
  year    = {2018},
  issn    = {1389-1286},
  doi     = {10.1016/j.comnet.2018.01.040},
  author  = {Andr{\'e}s F. Murillo Piedrahita and Vikram Gaur and Jairo Giraldo and Alvaro A. Cardenas and Sandra Julieta Rueda},
}

@article{related-work-heuristic,
author = {Babar, Ayesha and Halabi, Talal and Zulkernine, Mohammad},
title = {Autonomous and Adaptive Cyber Incident Detection and Response in Industrial Cyber-Physical Systems Using Hierarchical Reinforcement Learning},
year = {2026},
issue_date = {January 2026},
publisher = {Association for Computing Machinery},
address = {New York, NY, USA},
volume = {10},
number = {1},
issn = {2378-962X},
doi = {10.1145/3765622},
journal = {ACM Trans. Cyber-Phys. Syst.},
month = jan,
articleno = {8},
numpages = {27}
}

@article{related-work-heuristic2,
title = {An intrusion response approach based on multi-objective optimization and deep Q network for industrial control systems},
journal = {Expert Systems with Applications},
volume = {272},
pages = {126664},
year = {2025},
issn = {0957-4174},
doi = {https://doi.org/10.1016/j.eswa.2025.126664},
author = {Yiqun Yue and Dawei Zhao and Yang Zhou and Lijuan Xu and Yongwei Tang and Haipeng Peng},
}

@ARTICLE{related-work-heuristic3,
  author={Li, Xuan and Zhou, Chunjie and Tian, Yu-Chu and Qin, Yuanqing},
  journal={IEEE Transactions on Industrial Informatics}, 
  title={A Dynamic Decision-Making Approach for Intrusion Response in Industrial Control Systems}, 
  year={2019},
  volume={15},
  number={5},
  pages={2544-2554},
  doi={10.1109/TII.2018.2866445}
}

@misc{related-work-heuristic4,
      title={Multi-Agent Reinforcement Learning for Maritime Operational Technology Cyber Security}, 
      author={Alec Wilson and Ryan Menzies and Neela Morarji and David Foster and Marco Casassa Mont and Esin Turkbeyler and Lisa Gralewski},
      year={2024},
      eprint={2401.10149},
      archivePrefix={arXiv},
      primaryClass={cs.LG},
}

@ARTICLE{related-work-mdp,
  author={Chen, Hao and Lai, Yingxu and Liu, Jing and Wanyan, Hanxiao},
  journal={IEEE Transactions on Industrial Informatics}, 
  title={Interpretable Cross-Layer Intrusion Response System Based on Deep Reinforcement Learning for Industrial Control Systems}, 
  year={2024},
  volume={20},
  number={7},
  pages={9771-9781},
  doi={10.1109/TII.2024.3388672}
}

@INPROCEEDINGS{related-work-mdp2,
  author={Xu, Shoukun and Xie, Zihao and Zhu, Chenyang and Wang, Xueyuan and Shi, Lin},
  booktitle={2023 IEEE 29th International Conference on Parallel and Distributed Systems (ICPADS)}, 
  title={Enhancing Cybersecurity in Industrial Control System with Autonomous Defense Using Normalized Proximal Policy Optimization Model}, 
  year={2023},
  volume={},
  number={},
  pages={928-935},
  doi={10.1109/ICPADS60453.2023.00138}}

@ARTICLE{related-work-mdp3,
  author={Chen, Lin and Lai, Yingxu and Zhao, Peng and Xie, Baoshan and Zhang, Yan},
  journal={IEEE Transactions on Consumer Electronics}, 
  title={Safety-Aware Intrusion Response System Based on Safe Reinforcement Learning for Cyber-Physical Systems}, 
  year={2026},
  volume={72},
  number={2},
  pages={2711-2723},
  doi={10.1109/TCE.2026.3674866}}

@article{related-work-pomdp,
  title={Reinforcement Learning for Industrial Control Network Cyber Security Orchestration},
  author={John Mern and Kyle Beltran Hatch and Ryan Silva and Jeffrey S. Brush and Mykel J. Kochenderfer},
  journal={ArXiv},
  year={2021},
  volume={abs/2106.05332},
}

@ARTICLE{related-work-pomdp2,
  author={Lai, Yingxu and Zhao, Peng and Wang, Ziqi},
  journal={IEEE Internet of Things Journal}, 
  title={MFIR: Model-Free Intrusion Response for Partially Observable Industrial Control Systems}, 
  year={2026},
  volume={},
  number={},
  pages={1-1},
  doi={10.1109/JIOT.2026.3685344}}

@article{related-work-game,
title = {Bayesian and stochastic game joint approach for Cross-Layer optimal defensive Decision-Making in industrial Cyber-Physical systems},
journal = {Information Sciences},
volume = {662},
pages = {120216},
year = {2024},
issn = {0020-0255},
doi = {https://doi.org/10.1016/j.ins.2024.120216},
author = {Pengchao Yao and Zhengze Jiang and Bingjing Yan and Qiang Yang and Wenhai Wang},
}

@article{related-work-game2,
  author={Zhong, Kai and Yang, Zhibang and Xiao, Guoqing and Li, Xingpei and Yang, Wangdong and Li, Kenli},
  journal={IEEE Transactions on Parallel and Distributed Systems}, 
  title={An Efficient Parallel Reinforcement Learning Approach to Cross-Layer Defense Mechanism in Industrial Control Systems}, 
  year={2022},
  volume={33},
  number={11},
  pages={2979-2990},
  doi={10.1109/TPDS.2021.3135412}
}

@ARTICLE{related-work-it-pomdp,
  author={Miehling, Erik and Rasouli, Mohammad and Teneketzis, Demosthenis},
  journal={IEEE Transactions on Information Forensics and Security}, 
  title={A POMDP Approach to the Dynamic Defense of Large-Scale Cyber Networks}, 
  year={2018},
  volume={13},
  number={10},
  pages={2490-2505},
  doi={10.1109/TIFS.2018.2819967}}

@INPROCEEDINGS{related-work-it-pomdp2,
  author={Le, Duc Huy and Stadler, Rolf},
  booktitle={2025 21st International Conference on Network and Service Management (CNSM)}, 
  title={Learning Optimal Defender Strategies for CAGE-2 using a POMDP Model}, 
  year={2025},
  volume={},
  number={},
  pages={1-9},
  doi={10.23919/CNSM67658.2025.11297482}}

@ARTICLE{related-work-it-game,
  author={Zhang, Lefeng and Zhu, Tianqing and Hussain, Farookh Khadeer and Ye, Dayong and Zhou, Wanlei},
  journal={IEEE Transactions on Information Forensics and Security}, 
  title={A Game-Theoretic Method for Defending Against Advanced Persistent Threats in Cyber Systems}, 
  year={2023},
  volume={18},
  number={},
  pages={1349-1364},
  doi={10.1109/TIFS.2022.3229595}}

@INPROCEEDINGS{related-work-it-game2,
  author={Hammar, Kim and Stadler, Rolf},
  booktitle={2020 16th International Conference on Network and Service Management (CNSM)}, 
  title={Finding Effective Security Strategies through Reinforcement Learning and Self-Play}, 
  year={2020},
  volume={},
  number={},
  pages={1-9},
  doi={10.23919/CNSM50824.2020.9269092}}

@article{related-work-it-causal-model,
  title={Optimal defender strategies for CAGE-2 using causal modeling and tree search},
  author={Hammar, Kim and Dhir, Neil and Stadler, Rolf},
  journal={arXiv preprint arXiv:2407.11070},
  year={2024}
}

@INPROCEEDINGS{related-work-it-attack-graph,
  author={Nyberg, Jakob and Johnson, Pontus and M{\'e}hes, Andr{\'a}s},
  booktitle={2022 IEEE/IFIP Network Operations and Management Symposium (NOMS)}, 
  title={Cyber threat response using reinforcement learning in graph-based attack simulations}, 
  year={2022},
  volume={},
  number={},
  pages={1-4},
  doi={10.1109/NOMS54207.2022.9789835}}

@misc{related-work-it-llm,
      title={Incident Response Planning Using a Lightweight Large Language Model with Reduced Hallucination}, 
      author={Kim Hammar and Tansu Alpcan and Emil C. Lupu},
      year={2025},
      eprint={2508.05188},
      archivePrefix={arXiv},
      primaryClass={cs.CR},
}

@phdthesis{kim-thesis,
  author = {Hammar, K.},
  title  = {Optimal Security Response to Network Intrusions in IT Systems},
  school = {Kungliga Tekniska h{\"o}gskolan},
  address = {Stockholm},
  year   = {2024},
  type   = {PhD dissertation}
}

@misc{implement-github,
author = {Le, Duc Huy},
month = may,
title = {{Intrusion Response in OT systems (implementation)}},
version = {1.0},
year = {2026},
howpublished = {\url{https://github.com/duchuyle108/viper}},
}

@misc{docker,
author = {Merkel, Dirk},
title = {Docker: lightweight Linux containers for consistent development and deployment},
year = {2014},
issue_date = {March 2014},
publisher = {Belltown Media},
address = {Houston, TX},
volume = {2014},
number = {239},
issn = {1075-3583},
month = mar,
articleno = {2}
}

@misc{containerlab,
  author       = {{Nokia SR Linux Labs}},
  title        = {Containerlab},
  year         = {2026},
  url          = {https://containerlab.dev},
  note         = {Container-based networking lab orchestrator},
}

@article{openplc,
  author  = {Alves, Thiago and Morris, Thomas H.},
  title   = {{OpenPLC}: An {IEC} 61,131--3 compliant open source industrial controller for cyber security research},
  journal = {Computers \& Security},
  volume  = {78},
  pages   = {364--379},
  year    = {2018},
  doi     = {10.1016/j.cose.2018.07.007}
}

@manual{modbus,
  title        = {{MODBUS Messaging on TCP/IP Implementation Guide}},
  organization = {{Modbus Organization}},
  version      = {1.0b},
  year         = {2006},
  url          = {https://assets.noviams.com/novi-file-uploads/modbus/pdfs-and-documents/Modbus_Messaging_Implementation_Guide_V1_0b.pdf},
  note         = {Accessed: 2026-06-02}
}

@inproceedings{ovs-switch,
author = {Pfaff, Ben and Pettit, Justin and Koponen, Teemu and Jackson, Ethan J. and Zhou, Andy and Rajahalme, Jarno and Gross, Jesse and Wang, Alex and Stringer, Jonathan and Shelar, Pravin and Amidon, Keith and Casado, Mart\'{\i}n},
title = {The design and implementation of open vSwitch},
year = {2015},
isbn = {9781931971218},
publisher = {USENIX Association},
address = {USA},
booktitle = {Proceedings of the 12th USENIX Conference on Networked Systems Design and Implementation},
pages = {117–130},
numpages = {14},
location = {Oakland, CA},
series = {NSDI'15}
}

\end{document}